\documentclass[proceedings,nohyper]{JHEP}

\newcommand{\eqn}[1]{(\ref{#1})}
\newcommand{\complex}{{\bb C}} 
\newcommand{\complexs}{{\bbs C}} 
\newcommand{\zed}{{\bb Z}} 
\newcommand{\reals}{{\bbs R}} 
\newcommand{\zeds}{{\bbs Z}} 
\newcommand{\id}{{\bbs I}} 
\newcommand{\alg}{{\cal A}} 
\newcommand{\quater}{{\bbs H}} 
\newcommand{\NO}{\,\mbox{$\circ\atop\circ$}\,} 

\font\mybb=msbm10 at 8pt
\def\bb#1{\hbox{\mybb#1}}
\font\mybbs=msbm10 at 10pt
\def\bbs#1{\hbox{\mybbs#1}}

\def\nn{\nonumber}
\def\pvint{{\int\!\!\!\!\!\!-}}
\def\e{{\rm e}}
\def\semiprod{{\supset\!\!\!\!\!\!\!\times~}}

\def\beq{\begin{equation}}
\def\eeq{\end{equation}}
\def\bea{\begin{eqnarray}}
\def\eea{\end{eqnarray}}
\def\bd{\begin{displaymath}}
\def\ed{\end{displaymath}}

\conference{Noncommutative Geometry and String Duality}

\title{Noncommutative Geometry and String Duality\thanks{Based on talk given
by the first author at the 6th Hellenic School and Workshop on Elementary
Particle Physics, Corfu, Greece, September 6--26, 1998. To be published in {\it
JHEP Proceedings}.}}

\author{F. Lizzi$^a$ and R.J. Szabo$^b$\\
        $^a$ Dipartimento di Scienze Fisiche, Universit\`a di Napoli Federico
II\\
and INFN, Sezione di Napoli, Mostra d'Oltramare Pad. 20, 80125 Napoli, Italy\\
E-mail: \email{fedele.lizzi@na.infn.it}\\ \hfill \\ $^b$ The Niels Bohr
Institute\\ Blegdamsvej 17, DK-2100 Copenhagen \O, Denmark\\
E-mail: \email{szabo@nbi.dk}}

\abstract{A review of the applications of noncommutative geometry to a
systematic formulation of duality symmetries in string theory is presented. The
spectral triples associated with a lattice vertex operator algebra and the
corresponding Dirac-Ramond operators are constructed and shown to naturally
incorporate target space and discrete worldsheet dualities as isometries of the
noncommutative space. The target space duality and diffeomorphism symmetries
are shown to act as gauge transformations of the geometry. The connections with
the noncommutative torus and Matrix Theory compactifications are also
discussed. \\\hfill\\DSF/13-99  ,  NBI-HE-99-08  ,  hep-th/9904064 \hfill April
1999}

\begin{document}

\section{Introduction}

String theory in the previous decade raised some expectations about the
nature of geometry at very small distance scales. Because strings have a
finite intrinsic length scale $l_s$, it may not be possible to observe
distances smaller than $l_s$. Thus if one uses only strings as probes of
short distance structure, the conventional ideas of general relativity
break down at lengths of the order of $l_s$. This is exemplified through
string modified uncertainty relations \cite{ven} which yield an absolute
lower bound on the measurability of lengths in the spacetime. The spacetime
coordinates thus become smeared out and at short distances the notion of a
``point'' becomes meaningless.

Another piece of evidence is the $T$-duality symmetry of strings compactified
on a circle $S^1$ of radius $R$ \cite{duality}. This is a quantum symmetry
which maps the string theory onto one with target space the circle of dual
radius $\tilde R=l_s^2/R$, and at the same time interchanges the momenta of the
strings with their winding numbers around the $S^1$ in the spectrum of the
quantum string theory. Because of this symmetry, the moduli space of string
theories on $S^1$ is parametrized by radii $R\geq l_s$, and very small circles
are unobservable since the corresponding string theory can be mapped onto a
completely equivalent one living on a very large $S^1$. This leads to the
notion of quantum geometry, defined to be the appropriate modification of
classical
general relativity implied by string theory.

The uncertainty relations tell us that spacetime at very small distances
should be thought of as a quantum object. The appropriate mathematical
arena for the study of such ``pointless'' geometry is the theory of
algebras started by von Neumann, which in more modern times has developed
into noncommutative geometry \cite{connes}. Noncommutative geometry
presents an alternative, algebraic approach to the study of Riemannian
geometry and its generalizations, such as those hinted by string theory. In
this paper we shall discuss the applications of noncommutative geometry
towards a systematic development of the notion of quantum geometry.

Despite the implications on short distance structure presented by string
theory, many questions have been answered using only classical geometry. This
has occured in part because of the extremely rich mathematical structures
embedded into string compactifications which enables one to develop to a great
extent effective field theories on moduli spaces. However, the effective field
theories hide the true internal symmetries of string theory, and to study the
internal Kaluza-Klein spaces
the notion of spacetime geometry needs a drastic modification. This has become
especially clear over the last few years when it has been realized that the
low energy effective field theory for D-branes in string theory has
configuration space which is described in terms of non-commuting, matrix valued
spacetime coordinate fields \cite{wittenp}. This has led to, among other
things, the Matrix Theory conjecture \cite{bfss}, which proposes a light-cone
frame description of M-theory in terms of the Hamiltonian dynamics of
D0-branes. This rich structure indicates that some sort of generalization of
geometry is needed to describe the internal degrees of freedom implied by
D-branes and M-theory, and indeed it has been shown recently that
noncommutative geometry is the natural setting in which to study toroidal
compactifications of Matrix Theory \cite{cds}.

In this paper we shall review the formulation of quantum geometry through
the techniques of noncommutative geometry, based mostly on the approach
developed in \cite{fg}--\cite{size} that constructs a ``space'' in which
string duality is naturally realized as a true geometric symmetry. Starting
from a brief review of the ideas from noncommutative geometry that we shall
need, we shall describe the construction of the Fr\"ohlich-Gaw\c{e}dzki
geometry \cite{fg} which is based upon the algebraic properties of vertex
operator algebras. We will then describe how string duality naturally leads
to the quantum geometry of classical spacetimes within this framework, and
how the dualities manifest themselves as internal gauge symmetries of the
noncommutative geometry \cite{lscmp,lsplb,lscsf,ss}. This latter property
is formalized by a remarkable connection between string geometry and the
noncommutative torus \cite{lls}, which also allows us to relate this
worldsheet approach to the target space descriptions using Matrix Theory
compactifications.

\section{Spectral Triples in Noncommutative Geometry}

Noncommutative geometry is the study of geometric spaces (and their
generalizations) using algebras of fields defined on them. In this article we
shall discuss how to describe stringy spacetime as a noncommutative geometry,
and how the symmetries of the theory (such as $T$-duality and spacetime
diffeomorphisms) are realized as gauge transformations. The starting point is
to discuss an algebraic framework for ordinary {\it commutative} geometry.
Usually a compact Riemannian manifold $M$ is characterized as a topological
space on which
locally it is possible to introduce points $x\in M$ characterized by a finite
number of real numbers $x^i\in\reals$. Distances in $M$ are determined by the
metric of the space,
\beq
ds^2=g_{ij}(x)~dx^i\,dx^j
\label{ds2}\eeq
via the formula for geodesic length
\beq
d(x,y)=\inf_{\gamma_{x,y}}\,\mbox{$\int_{\gamma_{x,y}}$}\,ds
\label{dxy}\eeq
where $\gamma_{x,y}$ is a path from the point $x$ to the point $y$ in $M$.

There is a dual description of the topology and differentiable structure of a
smooth manifold which is provided by the $*$-algebra
$\alg=C^\infty(M,\complexs)$ of smooth complex-valued functions
$f:M\to\complexs$ (This algebra can be thought of as the algebra ``generated''
by the points of $M$). The completion of this algebra is the {\it commutative}
$C^*$-algebra $C^0(M,\complexs)$ of continuous complex-valued functions on $M$
with the $L^\infty$-norm
\beq
\|f\|_\infty=\sup_{x\in M}|f(x)|
\label{Linfnorm}\eeq
The algebra $C^0(M,\complexs)$ encodes all of the information about the
topology of the
space through the continuity criterion. Thus, in general, given a topological
space one may naturally associate to it an abelian $C^*$-algebra. That the
converse is also true is known as the Gel'fand-Naimark theorem \cite{fd}.
Namely, there is an isomorphism between the category of {\it Hausdorff}
topological spaces $M$ and the category of {\it commutative} $C^*$-algebras
$\alg$. The Gel'fand-Naimark functor is constructed by using the fact that
given an abelian $C^*$-algebra $\alg$, it is possible to reconstruct a
topological space $M$ as the structure space of characters of the algebra, i.e.
the $*$-linear multiplicative functionals $\chi:\alg\to\complexs$. Points $x\in
M$ are then obtained via the identification
\beq
\chi_x(f)=f(x)~~~~~~,~~~~~~\forall f\in\alg
\label{chardef}\eeq
and the topology is obtained in an unambiguous way from the notion of pointwise
convergence (in the usual topology of $\complexs$). Note that for a commutative
algebra a character is the same thing as an irreducible representation of
$\alg$.

What this all means is that the study of the properties of topological
spaces can be substituted by a purely algebraic description in terms of
abelian $C^*$-algebras. A {\em noncommutative space} is then obtained by
replacing $C^0(M,\complexs)$ by some non-abelian $C^*$-algebra. In that
case, not all irreducible representations of the algebra are
one-dimensional and the identification of ``points'' becomes ambiguous. But
as we shall see, the purely algebraic approach of noncommutative geometry
is particularly well-suited to describe the intrinsic symmetries of string
theory. Note that the $*$-algebra $C^\infty(M,\complexs)$ also encodes
all of the information about the differentiable structure of a manifold $M$
through the smoothness criterion. In what follows it will suffice to have a
description in terms of only a dense subalgebra of a given $C^*$-algebra.

The metric aspect and other geometrical properties are introduced into this
framework by using the fact that {\it any} $C^*$-algebra can be represented
faithfully and unitarily as a subalgebra of the algebra ${\cal B}({\cal
H})$ of bounded operators acting on some separable Hilbert space $\cal H$
\cite{fd}. In the following often we will not distinguish between  the
abstract algebra $\alg$ and its representation $\pi(\alg)$ (the norm on
$\alg$ is thus always understood as the operator norm and the
$*$-involution as Hermitian conjugation). An ``infinitesimal length
element'' is introduced by the relation
\beq
ds^{-1}=D
\label{dsD}\eeq
where $D$ is a (not necessarily bounded) operator on $\cal H$ which is called a
generalized Dirac operator \cite{connes} and which satisfies the following
properties:
\begin{itemize}
\item{$D=D^\dagger$ (this ensures positivity $ds^2\geq0$)}
\item{$[D,f]\in{\cal B}({\cal H})~~\forall f\in\alg$}
\item{$D$ has compact resolvent}
\end{itemize}
A metric space structure is then given by the Connes distance function on the
structure space of $\alg$,
\beq
d(x,y)=\sup_{f\in\alg\,:\,\|[D,f]\|\leq1}|\chi_x(f)-\chi_y(f)|
\label{ncdxy}\eeq

The generalized Dirac operator also enables one to introduce concepts from
ordinary differential geometry. Using $D$ one may define a representation of
abstract differential one-forms as
\beq
\pi_D(f\,dg)=f\,[D,g]
\label{piDdef}\eeq
and similarly one can introduce a representation of higher degree forms (In
that case a quotient space must be considered in order to eliminate the
so-called junk forms) \cite{connes}. Gauge theories are also readily
generalized to this setting. Generally, within this algebraic framework a
vector bundle
corresponds to a finitely-generated projective module over the algebra $\alg$
(the corresponding space of smooth sections when $\alg=C^\infty(M,\complexs)$).
However, in the following we shall consider, for simplicity, only the case of
trivial bundles. A gauge potential $A$ is then a one-form of the type
\eqn{piDdef} and a connection is provided via the Dirac operator through the
definition of a gauge covariant derivative
\beq
D_A=D+A
\label{DAdef}\eeq
The gauge group is defined as the group of unitary elements of the algebra,
\beq
{\cal U}(\alg)=\left\{u\in\alg~|~u^\dagger u=uu^\dagger=\id\right\}
\label{unitary}\eeq
and gauge transformations are the inner automorphisms of the algebra, i.e.
the maps $g_u:\alg\to\alg$ which act as conjugation by a unitary element
$u\in{\cal U}(\alg)$,
\beq
g_u(f)=ufu^\dagger
\label{gu}\eeq
The usual ingredients of a gauge theory on a manifold may then be summarized as
the following algebraic elements:
\begin{itemize}
\item{Connections: $A=\sum_nf_n\,[D,g_n]$, $f_n,g_n\in\alg$}
\item{Curvature: $F=[D,A]+A^2$}
\item{Bosonic Action: $\pvint~F^2$ (where $\pvint$ is a regularized trace, e.g.
the Dixmier trace \cite{connes})}
\item{Fermionic Action: $\pvint~\overline{\psi}\,D_A\psi$}
\end{itemize}

The set of three ingredients $(\alg,{\cal H},D)$, i.e. a $*$-algebra $\alg$ of
bounded operators acting on a separable Hilbert space $\cal H$ and a
generalized Dirac operator $D$ on $\cal H$, is called a spectral triple (or a
Dirac K-cycle). The
spectral triple encodes all the topological and geometrical information about a
Riemannian manifold. But notice that its construction is made with no reference
to any underlying space, and that it also applies to generic (not necessarily
commutative) algebras. We shall now describe a number of examples.

\subsection*{Spin Manifolds}

If $M$ is a compact spin manifold, we take
\bea
\alg&=&C^\infty(M,\complexs)\nn\\{\cal
H}&=&L^2(M,S)\nn\\D&=&i\,\gamma^i\,\nabla_i
\label{spinman}\eea
where $L^2(M,S)$ is the Hilbert space of square integrable spinors on $M$ and
the algebra $\alg$ acts diagonally on $\cal H$ by pointwise multiplication.
Here $\gamma^i$ are the usual Dirac matrices generating the Clifford algebra
$\{\gamma^i,\gamma^j\}=2g^{ij}$ of $M$ and $\nabla=d+\Gamma$ is the usual
covariant derivative constructed from the spin connection of $M$. The invariant
line element of $M$ is now represented as the free massless fermion propagator
on $M$ and the distance function \eqn{ncdxy} is
\beq
d(x,y)=\sup_{|\nabla f|\leq1}|f(x)-f(y)|
\label{dxyspin}\eeq
Note that the distance formula \eqn{dxyspin}, which is defined in terms of
complex-valued functions on $M$, is dual to the geodesic distance formula
\eqn{dxy}, which is defined in terms of arcs connecting $x$ to $y$ in $M$.
Notice also how the Riemannian geometry of $M$ is naturally encoded within the
definition of the Dirac operator.

The action functionals described above corresponding to a gauge theory
constructed from this spectral triple yield the usual one for electrodynamics
on the manifold $M$, with gauge group the unimodular loop group
$C^\infty(M,S^1)$ of $U(1)$ gauge transformations on $M$. Note that the
spectral triple \eqn{spinman} is that which naturally arises from quantizing
the free geodesic motion of a test particle on $M$. Then $\alg$ is the algebra
of observables, $\cal H$ is the Hilbert space of physical states, and the
Hamiltonian $H=-D^2$ is the
Laplace-Beltrami operator of $M$. Thus ordinary (commutative) spaces can be
thought of as those probed by quantum mechanical test particles. It is in this
way that noncommutative geometry may be thought of as ``quantum geometry''.

\subsection*{Morita Equivalence}

The natural extension of the previous example is
\bea
\alg&=&C^\infty(M,\complexs)\otimes M_N(\complexs)\nn\\{\cal
H}&=&L^2(M,S)^{\oplus N}\nn\\D&=&i\,\gamma^i\,\nabla_i\otimes I_N
\label{spinmanN}\eea
where $M_N(\complexs)$ is the algebra of $N\times N$ complex-valued matrices
and $I_N$ is the $N\times N$ identity matrix. The gauge group is now the group
of $U(N)$ gauge transformations $C^\infty(M,U(N))$ on $M$ and the spinors of
$\cal H$ transform in the vector representation of $U(N)$. Since the algebra
$\alg$ of matrix-valued functions on the manifold $M$ is noncommutative, the
Gel'fand-Naimark theorem does not apply to this case and it is not possible to
formally identify ``points'' of a space. In fact, there is an $N$ dimensional
sphere of pure states at each point (corresponding to the various unitary
equivalent representations) which can be thought of as an internal Kaluza-Klein
isospin
space with points connected by $U(N)$ gauge transformations. Nevertheless, it
is clear that the configuration space of the quantum theory corresponding to
\eqn{spinmanN} is still the manifold $M$. The choices of abelian subalgebras of
$M_N(\complexs)$ create $N$ copies of the same manifold connected by gauge
transformation.

The apparent paradox is resolved by noticing that the algebra $M_N(\complexs)$
has only one non-trivial irreducible representation as a $C^*$-algebra. Thus
the spectral triples \eqn{spinman} and \eqn{spinmanN} both determine the same
space. This phenomenon is captured formally by saying that the algebras
$C^\infty(M,\complexs)$ and $C^\infty(M,\complexs)\otimes M_N(\complexs)$ are
Morita equivalent. A $C^*$-algebra $\cal B$ is Morita equivalent to a
$C^*$-algebra $\alg$ if it is isomorphic to the algebra ${\rm End}_\alg^0({\cal
E})$ of compact endomorphisms of some $\alg$-module $\cal E$. This means that
the two algebras become isomorphic upon tensoring them with the algebra of
compact operators. Morita equivalent $C^*$-algebras have equivalent
representation theories. They therefore differ only in the structure of their
internal spaces (i.e. their gauge symmetries). The action functionals
corresponding to \eqn{spinmanN} yield the usual massless $U(N)$ gauge theory on
the manifold $M$.

\subsection*{The Two-sheeted Spacetime}

Let us now consider the spectral triple \cite{standard}
\bea
\alg&=&C^\infty(M,\complexs)\otimes\zeds_2\nn\\{\cal
H}&=&L^2(M,S)\otimes\zeds_2\nn\\D&=&\pmatrix{i\,\gamma^i\,\nabla_i&m\cr
m^*&i\,\gamma^i\,\nabla_i\cr}
\label{2sheet}\eea
where $m$ is a fermion mass. The bosonic action functional corresponding to
\eqn{2sheet} gives not only the usual Yang-Mills term, but also the Higgs
potential with its biquadratic form. This (commutative) space therefore gives a
geometrical origin for the Higgs mechanism associated with the spontaneous
breaking of the $U(1)\times U(1)$ gauge symmetry of $\alg$ down to $U(1)$
corresponding to the diagonal projection of $\alg$ onto
$C^\infty(M,\complexs)$.

\subsection*{The Standard Model}

The previous example can be generalized to the noncommutative geometry
\cite{standard}
\bea
\alg&=&C^\infty(M,\complexs)\otimes[\complexs\oplus\quater\oplus
M_3(\complexs)]\nn\\{\cal
H}&=&L^2(M,S)\otimes[\complexs
\oplus\complexs^2\oplus\complexs^3]
\nn\\D&=&\pmatrix{i\,
\gamma^i\,\nabla_i\otimes I_3&{\cal M}
\cr{\cal M}^\dagger&i\,\gamma^i\,\nabla_i\otimes I_3\cr}
\label{standard}\eea
where $\quater$ is the algebra of quaternions and $\cal M$ is the $3\times3$
fermion mass matrix. The unimodular group of the algebra $\alg$ is the familiar
gauge group $C^\infty(M,U(1)\times SU(2)\times SU(3))$ of the standard model
and $\cal H$ represents the physical Hilbert space of six generation fermions.
The action functional consists of the Yang-Mills action and the Higgs term. The
spectral triple \eqn{standard} thereby shows how electroweak and chromodynamic
degrees of freedom are induced by the geometry involving a discrete internal
Kaluza-Klein space.

\subsection*{The Noncommutative Torus}

The spectral triples we have considered thus far all have the property that
they contain an underlying ordinary geometry, i.e. their algebras have the form
$\alg=C^\infty(M,\complexs)\otimes\alg_F$ where $\alg_F$ is a finite
dimensional algebra. We now turn to a genuine example of a noncommutative
geometry, in which it is impossible to think of ``points'', that will turn out
to play a prominant role in the string theory applications.

Consider a $d$ dimensional torus $T_d=\reals^d/2\pi\Gamma$, where $\Gamma$ is a
Euclidean lattice of rank $d$ with bilinear form $g_{ij}$. Let $\omega^{ij}$ be
a real-valued antisymmetric matrix. We define an algebra $\alg^{(\omega)}$ with
generators $U_i$, $i=1,\dots,d$, and relations
\bea
U_iU_i^\dagger&=&U_i^\dagger U_i~=~\id\nn\\U_iU_j&=&\e^{2\pi
i\omega^{ij}}\,U_jU_i
\label{nctorusrels}\eea
A generic ``smooth'' element $f$ of the completion $\alg_\infty^{(\omega)}$ of
$\alg^{(\omega)}$ is a linear combination of monomials in the $U_i$,
\beq
f=\sum_{p\in\Gamma^*}f_p\,U_1^{p_1}U_2^{p_2}\cdots U_d^{p_d}
\label{smoothelts}\eeq
where $\Gamma^*$ is the dual lattice to $\Gamma$ and $f_p$ are elements of the
Schwartz space ${\cal S}(\Gamma^*)$ of sequences of rapid decrease. The
abstract algebra $\alg^{(\omega)}_\infty$ can be given a concrete
representation as a quantum deformation of the algebra
$C^\infty(T_d,\complexs)$ of functions on the torus. The product of two
functions $f,g\in C^\infty(T_d,\complexs)$ is now given by
\beq
(f\star_\omega g)(x)=\exp\left(i\pi\omega^{ij}\,\mbox{$\frac\partial{\partial
x^i}\,\frac\partial{\partial x'^j}$}\right)f(x)g(x')\Bigm|_{x'=x}
\label{defproduct}\eeq
which is just the usual rule for multiplying Weyl ordered symbols of quantum
mechanical operators. With the product \eqn{defproduct} the generators in
\eqn{nctorusrels} are just the usual basic plane waves
\beq
U_i=\e^{ix^i}
\label{planewaves}\eeq
and the expansion \eqn{smoothelts} can be thought of as a generalized Fourier
series expansion.

It follows that $\alg_\infty^{(0)}\cong C^\infty(T_d,\complexs)$. For
$\omega^{ij}\neq0$, the algebra $\alg_\infty^{(\omega)}$ represents the
quotient of the ordinary torus by the orbit of a free particle in it whose
velocity vector forms an angle $\omega^{ij}$ with respect to cycles $i$ and $j$
of $T_d$. When the $\omega^{ij}$ are all rational numbers, the algebra
$\alg_\infty^{(\omega)}$ is therefore Morita equivalent to
$C^\infty(T_d,\complexs)$. When the $\omega^{ij}$ are irrational, the orbits
are dense and ergodic and the resulting quotient space is not a conventional
Hausdorff manifold. In any of the cases, an invariant integration may be
introduced via the unique trace $\pvint:\alg_\infty^{(\omega)}\to\complexs$
which
is given by the classical average
\beq
\pvint\,f=\int_{T_d}\prod_{i=1}^d\frac{dx^i}{2\pi}~f(x)=f_0
\label{trace}\eeq
and a Dirac operator may be introduced via the natural set of linear
derivations $\Delta_i:\alg_\infty^{(\omega)}\to\alg_\infty^{(\omega)}$,
$i=1,\dots,d$, defined by the logarithmic derivatives
\beq
\Delta_i(U_j)=\delta_{ij}\,U_j
\label{linderiv}\eeq

One aspect of this example which will be particularly important for us is the
set of Morita equivalence classes of noncommutative tori \cite{riefsch}. From
\eqn{nctorusrels} it follows that the noncommutative tori with deformation
parameters $\omega^{ij}$ and $\omega^{ij}+\lambda^{ij}$, with $\lambda^{ij}$
any antisymmetric integer valued matrix, are the same. This symmetry is part of
a larger group $O(d,d;\zeds)$ which parametrizes the Morita equivalence
classes. It acts naturally on the deformation matrix $\omega$ as (upon
picking a suitable basis of $\reals^{d,d}$)
\bea
&\omega\to\omega^*=(A\omega+B)(C\omega+D)^{-1}~~~~~~,\nn\\&{\rm
with}~~\pmatrix{A&B\cr C&D\cr}\in O(d,d;\zeds)
\label{moritaom}\eea
where $A,B,C,D$ are $d\times d$ integer valued matrices which satisfy the
relations
\bea
A^\top C+C^\top A&=&0~=~B^\top D+D^\top B\nn\\A^\top D+C^\top B&=&I_d
\label{ABCDrels}\eea
Later on we will see
that for toroidally compactified string theory, Morita equivalence of the
corresponding noncommutative tori is precisely the same notion as target space
duality on the associated toroidal string background.

\section{Noncommutative String\\ Spacetimes}

We now seek some sort of algebra which describes the ``noncommutative
coordinates'' of spacetime as seen by strings. The structure should be such
that at very large distance scales (much larger than $l_s$), where the
strings effectively become point particles which are well described by
ordinary quantum field theory, we recover a usual (commutative) spacetime
manifold $M$ as described in section 2. The elegant proposal of Fr\"ohlich
and Gaw\c{e}dzki \cite{fg}, which as we will see naturally incorporates
duality as a gauge symmetry of the corresponding spectral triple, is to
take $\alg$ to be the vertex operator algebra of the underlying worldsheet
conformal field theory of the string theory. Vertex operators describe
interactions of strings and they operate on the string Hilbert space as
insertions on the worldsheet corresponding to the emission or absorption of
string states. They therefore form the appropriate noncommutative algebra
which describes the quantum geometry of the ``space'' of interacting
strings. In this section we shall construct the spectral triple associated
with a toroidally compactified bosonic string theory. The theory of vertex
operator algebras \cite{fgr,voa} has a
distinguished place in mathematics (having connections with the theory of
modular functions, the Monster sporadic group, etc.). We shall start by
giving a very brief general overview of the definition and properties of a
vertex operator algebra.

\subsection*{Vertex Operator Algebras}

Any conformal field theory naturally has associated to it two chiral algebras
${\cal E}^\pm$ which form the operator product algebras of the
(anti-)holomorphic fields of the theory. They contain two mutually commuting
representations of the infinite dimensional Virasoro algebra (generating the
conformal invariance of the theory)
\beq
\left[L_k^\pm,L_m^\pm\right]=(k-m)L_{k+m}^\pm+\mbox{$\frac
c{12}$}\,(k^3-k)\delta_{k+m,0}
\label{virasoro}\eeq
where $c$ is the central charge of the string theory. The algebras ${\cal
E}^\pm$ act densely on the Hilbert spaces ${\cal H}^\pm$, of left and right
handed string states respectively. The vertex operator algebra is
constructed using the operator-state correspondence of local quantum field
theory. Namely, to each state $\psi^{(\pm)}\in{\cal H}^\pm$ there
corresponds a chiral vertex operator $V_\pm(\psi^{(\pm)};z_\pm)$, where
$(z_+,z_-)$ are local coordinates on a Riemann surface (here we assume that the
worldsheet is the Riemann sphere). For a bosonic string theory, the vertex
operators can be expanded as Laurent series
\beq
V_\pm(\psi^{(\pm)};z_\pm)=\sum_{n\in\zed}\psi_n^{(\pm)}\,z_\pm^{-n-1}
\label{VLaurent}\eeq

The vertex operator algebra $\alg$ is now characterized by some algebraic
relations which can be summarized as follows. The commutation relations are
determined by the braiding relations
\bea
&
&\left[V_\pm(\psi_I^{(\pm)};z_\pm)\,
V_\pm(\psi_J^{(\pm)};w_\pm)\right]_{\gamma^\pm_{z,w}}\nn\\
& &=\sum_{K,L}\left(R^\pm\right)_{IJ}^{KL}~
V_\pm(\psi_K^{(\pm)};w_\pm)\,V_\pm(\psi_L^{(\pm)};z_\pm)\nn\\& &
\label{braid}\eea
where $R^\pm$ are called braiding matrices and $[\,\cdot\,]_{\gamma_{z,w}^\pm}$
is the operator which exchanges the two points $z_\pm\leftrightarrow w_\pm$
along a path $\gamma_{z,w}^\pm$ with (anti-)clockwise orientation on the
worldsheet. The products in \eqn{braid} are well defined operators on ${\cal
H}^\pm$ provided that $|z_\pm|<|w_\pm|$. Furthermore, chiral vertex operators
may be composed together using the fusion equations
\bea
&
&V_\pm(\psi_I^{(\pm)};z_\pm)\,V_\pm(\psi_J^{(\pm)};w_\pm)=\sum_{K,L}\left(F^\pm\right)_{IJ}^{KL}\nn\\& &\times\,V_\pm
\left(V_\pm(\psi_K^{(\pm)};z_\pm-w_\pm)\,\psi_L^{(\pm)};w_\pm\right)
\label{fusion}\eea
where $F^\pm$ are called fusion matrices. The braiding and fusion relations
completely characterize the chiral algebras ${\cal E}^\pm$. They can be
combined into a single relation known as the Jacobi identity of the vertex
operator algebra \cite{voa}, which can be thought of as a combination of the
classical Jacobi identity for Lie algebras and the Cauchy residue formula for
meromorphic functions.

The full, left-right symmetric vertex operator algebra is now obtained via the
sewing transformations
\bea
& &V(\psi_I;z_+,z_-)\nn\\& &=\sum_{K,L}D_I^{KL}~V_+(\psi_K^{(+)};z_+)\otimes
V_-(\psi_L^{(-)};z_-)\nn\\& &
\label{sewing}\eea
where $D_I^{KL}$ are complex valued sewing coefficients and
$\psi_I=\psi_I^{(+)}\otimes\psi_I^{(-)}$. The local fields \eqn{sewing} act
as operator valued distributions on the Hilbert space ${\cal
H}=\complexs^{\{D\}}\otimes{\cal H}^+\otimes{\cal H}^-$, where
$\complexs^{\{D\}}$ is the finite dimensional multiplicity space which
labels the various left-right sewings. The operator-state correspondence is
represented by the relation $V(\psi;0,0)|{\rm vac}\rangle=\psi\in{\cal H}$,
where $|{\rm vac}\rangle$ is the vacuum state of $\cal H$ (which we assume
is unique). Locality is the constraint that two operators of the type
\eqn{sewing} commute whenever their worldsheet arguments are space-like
separated. From this constraint it is possible to actually combine the
braiding and fusion relations into a reduced set of relations known as the
operator product expansion of two local conformal fields \eqn{sewing}. For
this, we grade the Hilbert space $\cal H$ by conformal dimensions
$\Delta_I^\pm$ which are the highest weights of the representations of the
Virasoro algebra in ${\cal E}^\pm$. The corresponding highest weight
vectors $\psi_I$ for ${\cal E}^+\otimes{\cal E}^-$ are called primary
states and are defined by
\beq
L_0^\pm\psi_I=\Delta_I^\pm\psi_I~~~~~~,~~~~~~L_k^\pm\psi_I=0~~\forall k>0
\label{highestwt}\eeq
The associated vertex operators are called primary fields and they satisfy the
differential equations
\bea
& &\left[L_k^\pm,V(\psi_I;z_+,z_-)\right]\nn\\&
&=\left(z_\pm^{k+1}\frac\partial{\partial z_\pm}+(k+1)\Delta_I^\pm
z_\pm^k\right)V(\psi_I;z_+,z_-)\nn\\& &
\label{primfield}\eea
With a suitable normalization of the two-point functions of primary fields of
fixed conformal dimension, one may derive the operator product expansion
\bea
& &V(\psi_I;z_+,z_-)\,V(\psi_J;w_+,w_-)\nn\\&
&=\sum_KC_{IJK}~(z_+-w_+)^{\Delta_K^+-\Delta_I^+-\Delta_J^+}\nn\\&
&~~~~\times(z_--w_-)^{\Delta_K^--\Delta_I^--\Delta_J^-}~V(\psi_K;z_+,z_-)\nn\\&
&
\label{ope}\eea
where the sum runs over a complete set of primary fields (equivalently
orthonormal primary states $\psi_K\in{\cal H}$), and $C_{IJK}$ are the constant
operator product expansion coefficients which are functions of the braiding,
fusion and sewing coefficients introduced above. The relation \eqn{ope}
completely characterizes the vertex operator algebra, which as we see is a
rather complicated unital $*$-algebra.

\subsection*{Lattice Vertex Operator Algebras}

We shall now specialize the above discussion to the case of closed bosonic
strings propagating in a $d$-dimensional toroidal target space
$T_d=\reals^d/2\pi\Gamma$. The classical string embedding fields in such a
target space are determined as the mod $2\pi\Gamma$ periodic solutions of the
two dimensional wave equation, which are given by the chiral multivalued
Fubini-Veneziano fields (in units with $l_s=1$)
\beq
X_\pm^i(z_\pm)=x_\pm^i+ig^{ij}p_j^\pm\log
z_\pm+\sum_{k\neq0}\frac1{ik}\,\alpha_k^{(\pm)i}\,z_\pm^{-k}
\label{fvfields}\eeq
where $g^{ij}$ is the matrix inverse of the metric $g_{ij}$ of $T_d$,
$(\alpha_k^{(\pm)i})^*=\alpha_{-k}^{(\pm)i}$ and $z_\pm=\e^{-i(\tau\pm\sigma)}$
with $(\tau,\sigma)$ local coordinates on the cylinder $\reals\times S^1$. The
left-right momenta are given by
\beq
p_i^\pm=\mbox{$\frac1{\sqrt2}$}\left(p_i\pm d_{ij}^\pm w^j\right)
\label{lrmom}\eeq
and the background matrices are
\beq
d_{ij}^\pm=g_{ij}\pm\beta_{ij}
\label{background}\eeq
with $\beta_{ij}$ the antisymmetric constant torsion form of the target space.
The zero modes $x^i=\frac1{\sqrt2}(x_+^i+x_-^i)\in T_d$ represent the position
of the center of mass of the string while $p_i\in\Gamma^*$ are the
corresponding momenta. The $\alpha$'s represent the vibrational modes of the
string and
$w^i\in\Gamma$ are the winding numbers which represent the number of times
that the string wraps around the cycles of the torus. The set of momenta
$(p^+,p^-)$ live in the even, self-dual Lorentzian lattice
$\Gamma^*\oplus\Gamma$.

Canonical quantization of this theory identifies the non-vanishing quantum
commutators
\bea
\left[p_i^\pm,x_\pm^j\right]&=&
i\,\delta_i^j\nn\\\left[\alpha_k^{(\pm)i},\alpha_m^{(\pm)j}\right]
&=&k\,g^{ij}\,\delta_{k+m,0}
\label{cancomms}\eea
The quantum fields \eqn{fvfields} therefore act on the Hilbert space
\beq
{\cal H}=L^2(T_d\times T_d^*,S)\otimes{\cal F}^+\otimes{\cal F}^-
\label{hilbert}\eeq
where $T_d^*=\reals^d/2\pi\Gamma^*$ is the dual torus to $T_d$ and $S\to
T_d\times T_d^*$ is the spin bundle over the double torus constructed from the
self-dual lattice $\Gamma^*\oplus\Gamma$. The $L^2$ space in \eqn{hilbert} is
constructed from the zero mode operators and is spanned by the plane wave
states
\beq
|q^+,q^-\rangle=\e^{-i(q_i-\beta_{ij}v^j)x^i-iv^ix^*_i}
\label{L2states}\eeq
where $x^*_i=\frac1{\sqrt2}g_{ij}(x_+^j+x_-^j)\in T_d^*$. The ${\cal
F}^\pm$ are bosonic Fock spaces generated by the oscillatory modes
$\alpha_k^{(\pm)i}$, respectively, and the unique vacuum state of \eqn{hilbert}
is $|{\rm vac}\rangle=|0,0\rangle\otimes|0\rangle_+\otimes|0\rangle_-$.

A vertex operator algebra for the toroidal compactification may now be
constructed using the operator-state correspondence, as described above. The
Hilbert space \eqn{hilbert} is spanned by states of the form
\bea
\psi&=&|q^+,q^-\rangle\otimes
\prod_{a=1}^Nr_i^{(a)+}\,\alpha_{-n_a}^{(+)i}
|0\rangle_+\nn\\& &\otimes\,\prod_{b=1}^Mr_j^{(b)-}\,\alpha_{-m_b}^{(-)j}
|0\rangle_-
\label{spanhilbert}\eea
where $(q^+,q^-),\,(r^{(a)+},r^{(a)-})\in\Gamma^*\oplus\Gamma$ and $n_a,m_b>0$.
To \eqn{spanhilbert} we associate the vertex operator
\bea
& &V(\psi;z_+,z_-)\nn\\& &=\NO
i\,V_{q^+q^-}(z_+,z_-)\,
\prod_{a=1}^N\frac{r_i^{(a)+}}{(n_a-1)!}\,
\frac{d^{n_a}X_+^i(z_+)}{dz_+^{n_a}}\nn\\&
&~~~~\times\prod_{b=1}^M\frac{r_j^{(b)-}}{(m_b-1)!}\,
\frac{d^{m_b}X_-^j(z_-)}{dz_-^{m_b}}\NO
\label{vertex}\eea
where $\NO\,\cdot\,\NO$ denotes the usual Wick normal ordering of operators,
and
\bea
& &V_{q^+q^-}(z_+,z_-)\nn\\&
&=(-1)^{q_iw^i}\,\NO\e^{-iq_i^+X_+^i(z_+)-iq_i^-X_-^i(z_-)}\NO\nn\\& &
\label{tachyon}\eea
are the basic, left-right symmetric tachyon vertex operators which generate a
unital $*$-algebra $\alg$ which acts diagonally on \eqn{hilbert}. The
operator-valued cocycle phases in \eqn{tachyon} are inserted to give the vertex
operators the correct locality relations. The algebraic properties of $\alg$
are described by the operator product expansion, which for the tachyon
generators reads
\bea
& &V_{q^+q^-}(z_+,z_-)\,V_{r^+r^-}(w_+,w_-)\nn\\& &=\e^{-\pi i\langle
q,r\rangle}~V_{r^+r^-}(w_+,w_-)\,V_{q^+q^-}(z_+,z_-)\nn\\& &
\label{opelattice}\eea
where we have assumed that $\pm\,{\rm arg}\,z_\pm>\pm\,{\rm arg}\,w_\pm$, and
\beq
\langle q,r\rangle=q_i^+g^{ij}r_j^+-q_i^-g^{ij}r_j^-
\label{naraininner}\eeq
is the bilinear form on $\Gamma^*\oplus\Gamma$. We shall discuss this
commutation relation in more detail later on. Note that the tachyon operators
create the states $|q^+,q^-\rangle\in L^2(T_d\times T_d^*,S)$ in which all
vibrational modes of the string are absent, i.e. they correspond to low energy
states of the spacetime, which as we will see lead to ordinary (commutative)
manifolds. On the other hand, the lowest symmetric stringy excitation of
commutative spacetime is the state
$|q^+,q^-\rangle\otimes\alpha_{-1}^{(+)i}|0\rangle_+\otimes
\alpha_{-1}^{(-)j}|0\rangle_-$ which is created by the graviton operator
\bea
& &V_{q^+q^-}^{ij}(z_+,z_-)\nn\\& &=\NO
i\,V_{q^+q^-}(z_+,z_-)\,\frac{dX_+^i(z_+)}{dz_+}\,
\frac{dX_-^j(z_-)}{dz_-}\NO\nn\\
\label{graviton}\eea
and represents the Fourier modes of the background matrices $d_{ij}^\pm$.

\subsection*{Spectral Triples for Toroidal Compactifications}

The final object we require to complete the string theory spectral triple
is an appropriate Dirac operator. In addition to the basic conformal
symmetry of the model (generated by reparametrizations of the worldsheet
coordinates), there is the target space reparametrization symmetry
$X_\pm^i(z_\pm)\to X_\pm^i(z_\pm)+\delta X_\pm^i(z_\pm)$, with $\delta
X_\pm^i(z_\pm)$ arbitrary periodic functions on $T_d$, which is generated
by the conserved currents
\beq
\delta_\pm^i(z_\pm)=-i\,z_\pm\,\frac{dX_\pm^i(z_\pm)}{dz_\pm}
\label{kacmoody}\eeq
The operators \eqn{kacmoody} generate a $u(1)_+^d\oplus u(1)_-^d$ Kac-Moody
algebra whose highest weight states are the tachyon vectors described above.
The spin bundle $S\to T_d\times T_d^*$ inherits a natural chirality grading
from the corresponding Clifford module over the double torus, which yields two
sets of Dirac matrices $\gamma_i^\pm$ that generate the Clifford algebra
$\{\gamma_i^\pm,\gamma_j^\pm\}=\pm2g_{ij}$ (with all other anticommutators
vanishing). This grading naturally splits the Hilbert space \eqn{hilbert} as
${\cal H}={\cal H}^+\oplus{\cal H}^-$ into the $\pm1$ eigenspaces of the
corresponding chirality operator.

The chiral structure of the theory now enables us to introduce {\it two}
independent Dirac operators for the noncommutative geometry by \cite{lscmp}
\beq
D^\pm(z_\pm)=\gamma_i^\pm\,\delta_\pm^i(z_\pm)=\sum_{k\in\zed}
\gamma_i^\pm\,\alpha_k^{(\pm)i}\,z_\pm^{-k}
\label{diracchiral}\eeq
where $\alpha_0^{(\pm)i}=g^{ij}p_j^\pm$. The choices \eqn{diracchiral} are not
as ad-hoc as they may first seem. First of all, the operator $\delta_\pm$
generates reparametrizations of the target space, so that the first equality in
\eqn{diracchiral} is just the usual relationship between the Dirac operator and
a covariant ``derivative''. Secondly, it is possible to show that the
energy-momentum tensor of the conformal field theory is related to the Dirac
operator by
\beq
T^\pm(z_\pm)=\sum_{k\in\zed}L_k^\pm\,z_\pm^{-k-2}=-\NO D^\pm(z_\pm)^2\NO
\label{enmom}\eeq
so that the square of the Dirac operator is related to the Hamiltonian in the
same way as in the case of a spin-manifold. It can therefore be thought as
generating the appropriate ``Laplace-Beltrami operator'' for the Riemannian
geometry. Finally, it can also be shown that \eqn{diracchiral} are the Ramond
sector fermionic zero modes of the $N=1$ worldsheet supercharges which generate
supersymmetry transformations of the appropriate supersymmetrization of the
underlying worldsheet conformal sigma model. The quantized fermionic zero modes
of these supercharges, when acting on states of vanishing total spacetime
momentum, generate the deRham complex of the manifold $T_d$ -- the two
operators can be identified with the exterior derivative and co-derivative,
while harmonic forms correspond to supersymmetric (physical) states. This is
just the classic Witten complex associated with two-dimensional $N=1$
supersymmetric sigma models \cite{witten}. Thus the Dirac operators
\eqn{diracchiral} are indeed appropriate to the quantum geometry associated
with the string theory. The existence of two independent Dirac operators means
that there are two spectral triples that may be constructed. We shall see that
this feature severely restricts the effective spacetime geometry and is
ultimately responsible for the occurence of duality symmetries in the quantum
geometry.

\section{Target Space Duality}

As we will now show, the existence of two natural Dirac operators for the
noncommutative geometry of the previous section is not an ambiguous property
and is directly tied with the notion of duality. The main feature is that there
are several isometries of the spectral triple that relate the chirally
symmetric and antisymmetric Dirac operators $D,\overline{D}=D^+\pm D^-$. An
isometry in the present context is a unitary operator $T:{\cal H}\to{\cal H}$
which is an automorphism of the vertex operator algebra $\alg$, i.e. $T\,\alg
\,T^{-1}=\alg$ (and in particular it preserves the operator-state
correspondence), and which relates the two Dirac operators via
\beq
D\,T=T\,\overline{D}
\label{Diso}\eeq
This implies that, at the level of their spectral triples, the two spacetimes
associated with $D$ and $\overline{D}$ are the same,
\beq
(\alg\,,\,{\cal H}\,,\,D)~\cong~(\alg\,,\,{\cal H}\,,\,\overline{D})
\label{speciso}\eeq
Since a change of Dirac operator in noncommutative geometry corresponds to a
change in metric on the ``manifold'', the isomorphism \eqn{speciso} is simply
the statement of general covariance of the noncommutative string spacetime.

A spacetime duality transformation is defined as an isomorphism of this
type which identifies subspaces of the two spectral triples in
\eqn{speciso} representing (classically) distinct ordinary spacetimes. The
idea is represented symbolically by the diagram:
\beq{\begin{array}{crc}
(\alg,{\cal H},D)&{\buildrel T\over\longrightarrow}&(\alg,{\cal
H},\overline{D})\\{\scriptstyle\Pi_0}\downarrow&~~&
\downarrow{\scriptstyle\overline{\Pi}_0}\\(\alg_0,{\cal H}_0,D_0)&~~&
(\overline{\alg}_0,\overline{\cal H}_0,\overline{D}_0)\end{array}}
\label{dualitydiag}\eeq
In this diagram, the top line represents the isomorphisms between the full
spectral triples, while $\Pi_0:{\cal H}\to{\cal H}_0$ and
$\overline{\Pi}_0:{\cal H}\to\overline{\cal H}_0$ are orthogonal projections
onto subspaces which represent classical spacetime geometries, e.g.
$(\alg_0,{\cal
H}_0,D_0)=(C^\infty(T_d,\complexs),L^2(T_d,S),ig^{ij}\gamma_i\partial_j)$
represents the ordinary torus $T_d$ with its flat metric $g_{ij}$. From the
point of view of ordinary geometry, there is no reason for the two classical
spacetimes in the bottom line of \eqn{dualitydiag} to be the same. However,
their embeddings into the full spectral triple representing the noncommutative
string spacetime defines an equivalence relation under the action of the
unitary isomorphism $T$ which identifies them at the level of the quantum
geometry, i.e. the mappings in \eqn{dualitydiag} do not commute with the
projection operators, and hence distinct classical spacetimes are identified.
This is a very natural and powerful way to characterize string geometry in
terms of different projections of the same spectral triple.

First, we shall establish that there indeed does exist a natural projection of
the noncommutative string spacetime onto a classical, low-energy sector.
Consider the subspace
\beq
\overline{\cal H}_0=\ker D\cong\bigotimes_{i=1}^d\left(\overline{\cal
H}_0^{(+)i}\oplus\overline{\cal H}^{(-)i}_0\right)
\label{barH0}\eeq
The states in $\overline{\cal H}_0$ are projected onto the vacuum sectors of
the Fock spaces, i.e. their oscillatory parts vanish, as anticipated for a
low-energy regime of the string theory. The space $\overline{\cal H}_0^{(+)i}$
consists of those states which have vanishing momentum $p_i=0$ and whose spinor
components carry the chiral action of the spin group defined by the action of
the Dirac matrices as $g^{jk}d_{ki}^+\gamma_j^+=g^{jk}d_{ki}^-\gamma_j^-$ (upon
choosing appropriate boundary conditions for the spinors with respect to a
homology basis of $T_d\times T_d^*$) in the $i$-th direction of the target
space. On the other hand, $\overline{\cal H}^{(-)i}_0$ consists of those states
which have zero winding number $w^i=0$ and which transform under the antichiral
spinor representation defined by $\gamma_i^+=-\gamma_i^-$.

The $2^d$ subspaces in the decomposition \eqn{barH0} are all naturally
isomorphic to one another \cite{lscmp} under ``partial'' $T$-duality
transformations $\overline{\cal H}^{(+)i}\leftrightarrow\overline{\cal
H}^{(-)i}$ for each $i$. This simply means that this decomposition is
independent of the choice of spin structure on the torus (or equivalently of
the choice of spinor boundary conditions along the elements of a homology
basis), as it should be, and it is a remarkable fact that this independence is
in itself an internal duality symmetry of the string theory. It suffices then
to consider only the completely antichiral subspace
\beq
\overline{\cal H}_0^{(-)}=\overline{\cal H}_0^{(-)1}\otimes\overline{\cal
H}_0^{(-)2}\otimes\cdots\otimes\overline{\cal H}_0^{(-)d}
\label{antichiralsubsp}\eeq
which consists of those states of $\cal H$ which have the quantum
representations $\gamma_i^+=-\gamma_i^-\equiv\gamma_i$ and $w^i=0$ for all
$i=1,\dots,d$. This is precisely what one expects for the low-energy
(particle-like) regime of the string theory, whereby the chiral structure
disappears and there are no non-local string modes that wind around the
compactified directions of the target space. In fact, since
$|p,p\rangle=\e^{-ip_ix^i}$, it follows that $\overline{\cal H}_0^{(-)}\cong
L^2(T_d,S^-)$, where $S^-\to T_d$ is the projection of the spin bundle $S\to
T_d\times T_d^*$ onto its antichiral representation. It is also straightforward
to see that the restriction of the antichiral Dirac operator to the subspace
\eqn{antichiralsubsp} is
\beq
\overline{D}\,\overline{\Pi}_0^{(-)}=
i\,g^{ij}\,\gamma_i\,\mbox{$\frac\partial{\partial x^j}$}
\label{barDproj}\eeq
where $\overline{\Pi}_0^{(-)}:{\cal H}\to\overline{\cal H}_0^{(-)}$ is the
corresponding orthogonal projection. Finally, using the operator-state
correspondence, the corresponding algebra is taken to be the projection of the
commutant of the chiral Dirac operator in $\alg$ (the ``restriction'' of $\alg$
to \eqn{antichiralsubsp}),
\beq
\overline{\alg}_0^{(-)}=\overline{\Pi}_0^{(-)}\left({\rm
End}_D\alg\right)\overline{\Pi}_0^{(-)}
\label{baralg0def}\eeq
where ${\rm End}_D\alg=\{V\in\alg\,|\,[D,V]=0\}$ determines the largest
subalgebra of $\alg$ which acts densely on \eqn{barH0}. The elements of
\eqn{baralg0def} are those vertex operators which have no oscillatory modes and
which create string states of identical left and right chiral momentum, i.e.
$V_{qq}\sim\e^{-iq_ix^i}$, from which it follows that
$\overline{\alg}_0^{(-)}\cong C^\infty(T_d,\complexs)$. The construction
described above thereby produces a subspace of the spectral triple $(\alg,{\cal
H},\overline{D})$,
\bea
& &\left(\overline{\alg}_0^{(-)}\,,\,\overline{\cal
H}_0^{(-)}\,,\,\overline{D}\,\overline{\Pi}_0^{(-)}\right)\nn\\&
&~~~~~~\cong\Bigl(C^\infty(T_d,\complexs)\,,\,L^2(T_d,S^-)\,,\,i\,
g^{ij}\,\gamma_i\,\mbox{$\frac\partial{\partial x^j}$}\Bigr)\nn\\& &
\label{barlowenergy}\eea
The spectral triple \eqn{barlowenergy} represents the ordinary Riemannian
geometry of the torus $T_d$ with metric $g_{ij}$, and thus the Dirac operator
$D$ naturally defines the appropriate low-energy projection of the full
noncommutative string spacetime via its zero-mode eigenspace $\ker
D\subset{\cal H}$ (or its dual version ${\rm End}_D\alg\subset\alg$).

The target space duality transformation \eqn{dualitydiag} follows from the
observation that one could have chosen $\overline{D}$ instead of $D$ in the
definition \eqn{barH0} and carried out an analogous low energy projection.
Doing so, we define
\beq
{\cal H}_0=\ker\overline{D}\cong\bigotimes_{i=1}^d\left({\cal
H}_0^{(+)i}\oplus{\cal H}^{(-)i}_0\right)
\label{H0}\eeq
where ${\cal H}_0^{(+)i}$ consists of states with $\gamma_i^+=\gamma_i^-$ and
$w^i=0$, while the states of ${\cal H}_0^{(-)i}$ carry the quantum
representation $g^{jk}d_{ki}^+\gamma_j^+=-g^{jk}d_{ki}^-\gamma_j^-$ and
$p_i=0$. Taking again the canonical choice
\beq
{\cal H}_0^{(-)}={\cal H}_0^{(-)1}\otimes{\cal
H}_0^{(-)2}\otimes\cdots\otimes{\cal H}_0^{(-)d}
\label{chiralsubsp}\eeq
in which
$g^{jk}d_{kl}^+g^{li}\gamma_j^+=-g^{jk}d_{kl}^-g^{li}
\gamma_j^-\equiv\gamma_*^i$ and $p_i=0$ for all $i=1,\dots,d$, we find the
isomorphism ${\cal H}_0^{(-)}\cong L^2(T_d^*,S^-)$ under the identification
$|d^+w,-d^-w\rangle\leftrightarrow\e^{-iw^ix_i^*}$.
 The projected chiral Dirac operator is
\beq
D\,\Pi_0^{(-)}=i\,\gamma_*^i\,\mbox{$\frac\partial{\partial x^*_i}$}
\label{Dproj}\eeq
where $\Pi_0^{(-)}:{\cal H}\to{\cal H}_0^{(-)}$. Note that the Dirac matrices
$\gamma_*^i$ generate the Clifford algebra with the dual metric
\beq
\tilde g^{ij}=(d^+)^{ik}\,g_{kl}\,(d^-)^{lj}
\label{dualmetric}\eeq
on $T_d^*$ which defines an inner product on the dual lattice $\Gamma^*$ (equal
to the matrix inverse of $g_{ij}$ when $\beta=0$). The low-energy subalgebra of
$\alg$ is now
\beq
\alg_0^{(-)}=\Pi_0^{(-)}\left({\rm End}_{\overline{D}}\alg\right)\Pi_0^{(-)}
\label{alg0def}\eeq
which consists of vertex operators of the form
$V_{d^+v,-d^-v}\sim\e^{-iv^ix_i^*}$, so that $\alg_0^{(-)}\cong
C^\infty(T_d^*,\complexs)$. In this way we arrive at another low-energy
commutative subspace
\bea
& &\left(\alg_0^{(-)}\,,\,{\cal H}_0^{(-)}\,,\,D\,\Pi_0^{(-)}\right)\nn\\&
&~~~~~~\cong\Bigl(C^\infty(T_d^*,\complexs)\,,\,L^2(T_d^*,S^-)\,,\,i\,
\gamma_*^i\,\mbox{$\frac\partial{\partial x^*_i}$}\Bigr)\nn\\& &
\label{lowenergy}\eea
which represents the dual $d$-torus $T_d^*$ with metric $\tilde g^{ij}$.

{}From the point of view of classical general relativity, the two spacetimes
\eqn{barlowenergy} and \eqn{lowenergy} are inequivalent. However, consider the
unitary transformation $T=T_S\otimes T_X:{\cal H}\to{\cal H}$ which acts as the
gauge transformation
\bea
T_X&=&\e^{i{\cal G}_X}\in{\cal U}(\alg)\nn\\{\cal
G}_X&=&\mbox{$\frac\pi{2i}$}\left(J_+^+J_+^--J_-^+J_-^-\right)
\nn\\J_\pm^a(z_a)&=&\NO\e^{\pm ik_iX_a^i(z_a)}\NO
\label{Tdualtransf}\eea
where $a=\pm$ and $k_i$ is a Killing vector of $T_d$. It acts on the spectral
data of the full noncommutative geometry as \cite{lscmp}
\bea
T_X|p^+,p^-\rangle&=&(-1)^{p_iw^i}\nn\\&
&\times\,|(d^+)^{-1}p^+,-(d^-)^{-1}p^-\rangle\nn\\&
&\label{TXstates}\\T_X\,\alpha_n^{(\pm)i}\,T_X^{-1}&
=&\pm\,g_{jk}\,(d^\mp)^{ij}\,\alpha_n^{(\pm)k}\label{TXosc}
\\T_S\,\gamma_i^\pm\,T_S^{-1}&=&g^{jk}\,d_{ij}^\mp\,
\gamma_k^\pm\label{TSgamma}\\T_X\,V_{q^+q^-}\,T_X^{-1}
&=&(-1)^{q_iw^i}\,V_{q^+(d^+)^{-1},-q^
-(d^-)^{-1}}\nn\\& &
\label{TXV}\eea
Note that \eqn{TSgamma} implies that the metric $g_{ij}$ is mapped to its dual
\eqn{dualmetric}. It is straightforward to see that $T$ interchanges the two
Dirac operators as prescribed in \eqn{Diso}, and therefore gives the required
isomorphism of the corresponding spectral triples. In this way, the distinct
low energy classical spacetimes \eqn{barlowenergy} and \eqn{lowenergy} are
identified, leading to the celebrated $T$-duality transformation of toroidally
compactified string theory \cite{duality}, in which $d^\pm\to(d^\pm)^{-1}$,
$g\to\tilde g$, and $p_i\leftrightarrow w^i$. We see therefore that in the
framework of noncommutative geometry, the $T$-duality transformation of a
toroidal target space is just the low energy projection of a gauge
transformation on the noncommutative string spacetime.

The $T$-duality mapping actually determines only a $\zeds_2$ subgroup of a
larger discrete duality group of the string theory. The remaining generators
come from the web of transformations between the other various subspaces of
\eqn{barH0} and \eqn{H0}. These are all constructed explicitly in \cite{lscmp}.
For example, the mapping between the antichiral subspace \eqn{antichiralsubsp}
and the corresponding chiral subspace ${\cal H}_0^{(+)}$ determines the
worldsheet parity symmetry of the string theory, which acts on the worldsheet
by
interchanging the chiral structures and on the background data as
$\beta\to-\beta,d^\pm\to d^\mp$. Similarly, the mapping between
\eqn{antichiralsubsp} and the subspace of $\ker\overline{D}$ in which all
tensor components have chiral conditions except for the $i$-th one leads to the
factorized duality transformation of $T_d$, which can be thought of as the
$R\to1/R$ circle duality along the $i$-th cycle of $T_d$. In the present
framework, this transformation is also accompanied by a worldsheet parity map
in all of the other $d-1$ directions. When $d$ is even, the factorized
dualities lead to the phenomenon of mirror symmetry (and hence of spacetime
topology change) which exchanges the complex and K\"ahler structures of the
torus. In addition there other dualities which act trivially on the spectral
triples, but nonetheless do lead to non-trivial quantum dynamics and are
therefore considered as symmetries of the quantum geometry. These are the
changes of basis of the compactification lattice $\Gamma$, and the torsion
cohomology shift $\beta_{ij}\to\beta_{ij}+\lambda_{ij}$, with $\lambda_{ij}$ an
antisymmetric integer-valued matrix, which can be absorbed into a redefinition
of the momenta
$p_i\to p_i-\lambda_{ij}w^j$. Altogether these transformations generate the
discrete duality group which is the semi-direct product
\beq
G_d=O(d,d;\zeds)\,\semiprod\zeds_2
\label{dualitygp}\eeq
of the isometry group ${\rm Aut}(\Gamma^*\oplus\Gamma)=O(d,d;\zeds)$ by the
natural action of the worldsheet parity group $\zeds_2$. The duality group
\eqn{dualitygp} labels a large set of low energy theories and connects various
inequivalent classical spacetimes. Its action also leaves the spectra of the
Dirac operators $D$ and $\overline{D}$ invariant.

\section{Gauge Symmetries}

In the previous section we saw that the duality group \eqn{dualitygp} of
toroidally compactified string theory arises in a very natural way through the
automorphisms of the noncommutative string spacetime. With the exception of
worldsheet parity, the transformations we described above were all realized as
inner automorphisms and so represented gauge symmetries of the noncommutative
geometry \cite{lscmp}. On the other hand, worldsheet parity exchanges the left
and right chiral algebras ${\cal E}^\pm\to{\cal E}^\mp$ and therefore
represents an outer automorphism of the vertex operator algebra (since no
element of $\alg$ can accomplish this chirality reversal). In this final
section we shall study in more depth the automorphism group of the vertex
operator algebra, which represents the homeomorphisms of the noncommutative
space, and focus on the consequences of the realization of duality
transformations as gauge symmetries. Here we shall set $\beta_{ij}=0$.

\subsection*{Automorphisms of Lattice Vertex Operator Algebras}

Generally, the inner automorphisms of a $*$-algebra $\alg$ form a normal
subgroup ${\rm Inn}(\alg)$ of the automorphism group ${\rm Aut}(\alg)$. The
remaining symmetries ${\rm Out}(\alg)={\rm Aut}(\alg)/{\rm Inn}(\alg)$ form the
outer automorphisms of the algebra such that the automorphism group is the
semi-direct product
\beq
{\rm Aut}(\alg)={\rm Inn}(\alg)\,\semiprod{\rm Out}(\alg)
\label{autgp}\eeq
of ${\rm Inn}(\alg)$ by the natural action of ${\rm Out}(\alg)$. The inner
automorphisms represent internal fluctuations of the noncommutative geometry
corresponding to rotations \eqn{gu} of the elements of $\alg$. Consider the
example of a spin manifold discussed in section 2. In this case the inner
automorphisms of $\alg=C^\infty(M,\complexs)$ are all trivial (there are no
internal symmetries for a commutative algebra), while ${\rm Out}(\alg)\cong{\rm
Diff}(M)$ (given a diffeomorphism $\phi\in{\rm Diff}(M)$, one may construct the
natural automorphism $g_\phi(f)=f\circ\phi^{-1}~~\forall f\in\alg$). The outer
automorphisms in this case generate the general covariance of the space. On the
other hand, for the case $\alg=C^\infty(M,\complexs)\otimes M_N(\complexs)$,
while the outer automorphisms still generate general coordinate transformations
of the manifold $M$, the inner automorphisms generate the group of $U(N)$ gauge
transformations on $M$.

In these two examples there is a clear distinction between (internal) gauge
symmetries and outer automorphisms, again because there is an underlying
commutative algebra which is simply augmented by a discrete internal space.
This is not the case of the string theory spectral triple, which represents a
genuine noncommutative space. As a dramatic example, we recall that the
conserved currents \eqn{kacmoody} are the generators of target space
reparametrizations, so that a general coordinate transformation $X\to\xi(X)$ of
the spacetime may be represented by the gauge transformation \cite{lscmp}
\bea
T_\xi&=&\e^{i{\cal G}_\xi}\in{\cal U}(\alg)\nn\\{\cal
G}_\xi&=&\xi_i(X)\left(\delta_+^i+\delta_-^i\right)
\label{gencoordtransfs}\eea
However, these inner automorphisms are mapped onto outer automorphisms of the
commutative algebra $\overline{\alg}_0^{(-)}\cong C^\infty(T_d,\complexs)$,
representing the diffeomorphisms of the torus, under the low energy projection
onto the ordinary geometry of $T_d$. This is a very nontrivial fact, as it
implies that general covariance is manifested as a gauge symmetry of the full
noncommutative string spacetime, and in this sense general relativity is in
fact really a gauge theory. This is of course what one would like of a theory
that unifies all of the fundamental interactions, in that it puts gauge and
gravitational interactions on the same footing. Note that in this sense the
worldsheet parity transformations are implemented as ``diffeomorphisms'' of the
noncommutative string spacetime.

The realization of dualities as gauge symmetries yields a natural explanation
of the long standing puzzle as to the origin of string duality as part of some
mysterious gauge group \cite{duality}. In the present case, the conserved
currents \eqn{kacmoody} along with the chiral tachyon vertex operators generate
the affine Lie group ${\widehat{\rm Inn}}^{(0)}(\alg)$ of primary fields of
weight 1 \cite{lscsf}. The corresponding inner automorphisms give internal
perturbations which preserve the grading by conformal dimension and thereby
yield isomorphic conformal field theories. Generically, this group is just the
$U(1)_+^d\times U(1)_-^d$ Kac-Moody group, but there are points in the moduli
space of toroidal compactifications at which this group is enhanced to a
non-abelian gauge symmetry, such as an affine $SU(2)_+^d\times SU(2)_-^d$
Kac-Moody group. A natural subgroup of the automorphism group of a lattice
vertex
operator algebra is therefore given by \cite{lscsf}
\beq
{\rm Aut}^{(0)}(\alg)={\widehat{\rm Inn}}^{(0)}(\alg)\,\semiprod{\rm Out}(\alg)
\label{aut0alg}\eeq
where the group of outer automorphisms is
\beq
{\rm Out}(\alg)\cong O(d,d;\zeds)\,\semiprod O(2;\reals)
\label{outeralg}\eeq
where $O(d,d;\zeds)$ represents those automorphisms which preserve the bilinear
form of the Lorentzian lattice $\Gamma^*\oplus\Gamma$, while $O(2;\reals)$ is a
worldsheet symmetry group which is an extension of worldsheet parity that acts
by rotating the two chiral sectors of the theory among each other. The full
automorphism group is very large and is not known in explicit form. It
is related to some exotic mathematical constructions, such as the Monster
sporadic group \cite{voa}. But the qualitative form of \eqn{aut0alg} and the
occurence of duality as a gauge symmetry is a generic property of all duality
symmetries. For instance, all of these structures can be shown to hold for
asymmetric duality groups, such as those associated with heterotic strings
\cite{ss}, and also for non-perturbative $SL(2;\zeds)$ $S$-duality symmetries
\cite{lsplb}.

\subsection*{Duality Equivalence Classes}

In the previous section we saw the effects of quantum geometry at the level of
classical spectral triples. The full-blown noncommutative geometry comes into
play at the level of the automorphisms which implement these quantum
symmetries. For instance, the diffeomorphism gauge symmetries
\eqn{gencoordtransfs} are associated with the graviton operators \eqn{graviton}
which represent the lowest oscillatory excitation of the classical spacetime.
The other set of generators of ${\widehat{\rm Inn}}^{(0)}(\alg)$, i.e. the
tachyon vertex operators, turn out to also give a noncommutative perturbation
of classical spacetime that we shall now proceed to describe.

The operator product of two tachyon operators may be written as
\bea
& &V_{q^+q^-}(z_+,z_-)\,V_{r^+r^-}(w_+,w_-)\nn\\&
&=(z_+-w_+)^{q_i^+g^{ij}r_j^+}\,(z_--w_-)^{q_i^-g^{ij}r_j^-}\nn\\&
&~~~~~~\times\NO V_{q^+q^-}(z_+,z_-)\,V_{r^+r^-}(w_+,w_-)\NO\nn\\& &
\label{tachyonope}\eea
The expression \eqn{tachyonope} is in general singular at coinciding points of
the operators, and to make proper sense of it the vertex operators must be
suitably regularized. Doing so (see \cite{lls} for details), by interchanging
the order of the two operators the relation
\eqn{tachyonope} induces a local cocycle relation which, up to a permutation of
the coordinate directions of $T_d\times T_d^*$, is independent of the
worldsheet coordinates $z_\pm$ and $w_\pm$. This local cocycle relation may be
easily recognized as the defining relation of the noncommutative torus
\eqn{nctorusrels} with the generators $U_i$ identified with the tachyon
operators $V_{e^i_+e^i_-}$, where $e_\pm^i$ is the canonical basis of the
Lorentzian lattice
$\Gamma^*\oplus\Gamma$. The deformation parameter is related to the metric of
$T_d$ by \cite{lls}
\beq
\omega^{ij}={\rm sgn}(j-i)\,g^{ij}~~~~~~,~~~~~~i\neq j
\label{voadefpar}\eeq
More precisely, the tachyon algebra generated by the operators $V_{q^+q^-}$
defines a unitary equivalence class of self-dual $\zeds_2$-twisted projective
modules of rank $2d$ over the double noncommutative torus
$T_\omega^{(+)d}\times T_\omega^{(-)d}$ (one copy for each chiral sector of the
vertex operator algebra along with a $\zeds_2$ twist coming from the cocycle
factors in \eqn{tachyon}). Notice, however, that according to \eqn{opelattice}
the product of two projected tachyon operators (made via the operator-state
correspondence by mapping onto $L^2(T_d\times T_d^*,S)$) is commutative and
just represents the algebra of functions on $T_d\times T_d^*$. In the case at
hand, two given operators are first multiplied together in the full algebra,
and then the result is projected onto the tachyon sector of the noncommutative
geometry. Namely, given $V_0=\Pi_0\,V\,\Pi_0$ and $W_0=\Pi_0\,W\,\Pi_0$, where
$V,W\in\alg$ and $\Pi_0:{\cal H}\to L^2(T_d\times T_d^*,S)$, the deformed
product is given by
\beq
V_0\star_\omega W_0=\Pi_0\,(VW)\,\Pi_0
\label{voadefprod}\eeq
Thus the algebraic properties of the vertex operator algebra give a very
natural way to deform the algebra of functions on $T_d\times T_d^*$ which takes
into account of the internal oscillatory modes of the strings.

We see therefore that the intermediate regime separating the classical torus
from the highly noncommutative string spacetime with its oscillators turned on
is the tachyon algebra which is a module for the noncommutative torus. Note
that for very large distance scales, i.e. $g_{ij}\to\infty$, the deformation
parameter \eqn{voadefpar} vanishes and the tachyon algebra coincides with the
algebra of functions on the double torus. This is just the statement that at
very large distance scales we recover a classical spacetime. On the other hand,
at very small distance scales $g_{ij}\to0$ and thus, within the framework of
toroidally compactified string theory, spacetime at very short distances is a
noncommutative torus.

Having identified this intermediate regime allows us to explore features of the
noncommutative geometry without having to deal with all of the complicated
structures that make up the full vertex operator algebra. Moreover, it enables
us to put the notion of string duality into a more familiar mathematical
parlance. Namely, the deformation parameter \eqn{voadefpar} is the natural
induced antisymmetric bilinear form on the lattice $\Gamma$, and so it
transforms under the action of the target space duality group $O(d,d;\zeds)$ as
in \eqn{moritaom} \cite{duality,lscmp}. Each such orbit is implemented in the
full algebra $\alg$ by a gauge transformation $T$ as described above. Under the
equivalence relation generated by the isomorphism of full spectral triples, the
modules for the noncommutative tori with deformation parameters $\omega$ and
$\omega^*$ are identified. In other words, target space dualities are naturally
realized as the Morita equivalences among noncommutative tori. This remarkable
feature of string geometry can be thought of as lending a physical
interpretation to the mathematical notion of Morita equivalence in
noncommutative geometry. It is also in accord with the connections that exist
between the noncommutative torus and Matrix Theory \cite{cds}, in which the
deformation parameter $\omega$ is given by the light-like component of the
three-form tensor field coming from the corresponding reduction of
11-dimensional supergravity. In that case Morita equivalence relates dual
quantum theories which have the same BPS spectra \cite{schwarz}.

\subsection*{Universal Gauge Symmetry}

Let us now take the Lorentzian lattice to be
$\Gamma^*\oplus\Gamma=\zeds\oplus\omega\zeds$ and assume that the
deformation parameter is a rational number
\beq
\omega=M/N
\label{omegarat}\eeq
where $M$ and $N$ are relatively prime positive integers.
The center of $\alg^{(\omega)}$
\beq
{\cal Z}(\alg^{(\omega)})={\rm
span}_\complex\left\{U_1^{mN}U_2^{nN}~\Bigm|~m,n\in\zeds\right\}
\label{centeraomega}\eeq
is infinite dimensional, and the quotient
$\hat\alg^{(\omega)}=\alg^{(\omega)}/{\cal I}_\omega$ by the ideal ${\cal
I}_\omega$ generated by ${\cal Z}(\alg^{(\omega)})-\{\id\}$ is isomorphic, as a
unital $*$-algebra, to $M_N(\complexs)$. The clock algebra $U_1U_2=\e^{2\pi
i\omega}U_2U_1$ can be represented by the $N\times N$ clock and shift matrices
\bea
U_1&=&\pmatrix{1&0&0&\dots&0\cr0&q&0&\dots&0\cr0&0&q^2&\dots&0\cr\vdots&\vdots&
\vdots&\ddots&\vdots\cr0&0&0&\dots&q^{N-1}\cr}\nn\\
U_2&=&\pmatrix{0&1&0&\dots&0\cr0&0&1&\dots&0\cr\vdots&
\vdots&\vdots&\ddots&\vdots\cr0&0&0&\dots&1\cr1&0&0&\dots&0\cr}
\label{UVmatrix}\eea
where $q=\e^{2\pi iM/N}$.

The physical relevance of this special case is many-fold. First of all, any
``smooth'' element of the $su(N)$  Lie algebra can be expanded in terms of
products of the generators \eqn{UVmatrix}, such that in the limit
$N\to\infty,\omega\to0$ it becomes the Poisson-Lie algebra of smooth
functions on the 2-torus with respect to the usual Poisson bracket
\cite{bfss,fgr}. This is similar to what occured for the vertex operator
algebra, whereby some of the gauge symmetries of the noncommutative
geometry projected onto outer automorphisms of the classical spacetime. The
large-$N$ limit of $SU(N)$ is relevant to the Matrix Theory description of
M-theory, in which $N$ is proportional to the longitudinal momentum and the
limit $N\to\infty$ describes dynamics in the infinite momentum frame
\cite{bfss}. The gauge symmetry here is then represented by the infinite
unitary group $SU(\infty)$ which is an example of a universal gauge group
\cite{lscsf,Rajeev}. The canonical universal gauge groups of vertex
operator algebras, representing the algebras overlying all of the dynamical
symmetries of string theory, may thus be related, through the connection
with the noncommutative torus, to $SU(\infty)$.

More ambitiously, one can imagine that the noncommutative 2-torus with
rational deformation parameter \eqn{omegarat} is some sort of finite
dimensional approximation to the vertex operator algebra. In the correlated
limit $N,M\to\infty$ with $\omega$ a fixed irrational number, we recover the
genuine noncommutativetorus (the center \eqn{centeraomega} is then trivial)
which is related to a corresponding vertex operator algebra as described
above with an explicit representation of the universal gauge symmetry. The
noncommutative geometry associated with rank 2 lattice vertex operator
algebras may in this way be relevant to the true noncommutative geometric
structure encompassing the large-$N$ limit of Matrix Theory. However, in
order for this to be true, one needs to realize the noncommutative string
spacetime as some limit of a tower of finite dimensional matrix geometries.

The potential relevance to Matrix Theory compactifications comes from the
observation that the Morita equivalence of the noncommutative 2-torus can be
naturally interpreted as the zero area limit of the usual string theoretical
$SL(2;\zeds)$ duality in the presence of a background Neveu-Schwarz two-form
field $\beta$ \cite{dh}. In supergravity, this is not a geometrical symmetry in
the usual sense, but in the framework of noncommutative geometry it is. In the
present case, the field $\beta$ is given explicitly by \eqn{voadefpar} and in
the zero area limit we obtain a highly noncommutative structure. This is in
precise agreement with the idea that toroidal compactifications at very small
distance scales are equivalent to noncommutative tori.

\subsection*{Noncommutative Gauge Theory}

The mapping onto the noncommutative torus can also be used to give a dynamical
model for the duality symmetries of the noncommutative string spacetime,
through the construction of an explicitly duality symmetric noncommutative
gauge theory \cite{lls}. The action functional of the gauge theory defined from
the twisted module over the noncommutative torus is obtained using the
invariant integration \eqn{trace} and the derivations \eqn{linderiv}.
The interaction vertices are constructed from the commutator which is
represented by the Moyal bracket
\beq
[f,g]=\{f,g\}_\omega=f\star_\omega g-g\star_\omega f
\label{moyal}\eeq
where the deformed product of two functions $f,g\in C^\infty(T_d\times
T_d^*,\complexs)$ is given by
\bea
& &(f\star_\omega g)(x,x^*)\nn\\&
&=\exp\left[i\pi\omega^{ij}\left(\mbox{$\frac\partial{\partial
x^i}\,\frac\partial{\partial
x'^j}$}-g_{ik}g_{jl}\,\mbox{$\frac\partial{\partial
x_k^*}\,\frac\partial{\partial x^{\prime*}_l}$}\right)\right]\nn\\&
&~~~~~~\times\,f(x,x^*)\,g(x',x^{\prime*})
\Bigm|_{(x',x^{\prime*})=(x,x^*)}\nn\\& &
\label{defprodncgauge}\eea
Using the projections of the two Dirac operators $D$ and $\overline{D}$
introduced earlier onto the tachyon algebra, one may naturally write down
bosonic and fermionic Lagrangians in the usual spirit of noncommutative
geometry \cite{lls},
\bea
&&{\cal L}=\left(F+{ }^\star F\right)_{ij}\left(F+{ }^\star
F\right)^{ij}\nn\\ &&-i\overline{\psi_*}\gamma^i\left(\partial_i+
i{\buildrel\leftrightarrow\over
{A_i}}\right)\,\psi-i\overline{\psi}\gamma^*_i\left(\partial_*^i+i
{\buildrel\leftrightarrow\over{A_*^i}}\right)\,\psi_*\nn\\
\label{dualsymmaction}\eea
 The bosonic part yields a symmetrized
Yang-Mills type functional with the above interaction vertices, constructed
from two vector potentials $A(x,x^*)$ and $A_*(x,x^*)$. The fermionic part
contains fermion fields minimally coupled to the gauge potentials, again
determined by a non-local interaction vertex coming from the representation
of the noncommutative tachyon algebra on the Hilbert space $L^2(T_d\times
T_d^*,S)$, i.e.
\bea
(V_{q^+q^-}\,f)_{r^+r^-}&=&(-1)^{q_iv^i}~\e^{2\pi
i(q_i^+g^{ij}r_j^++q_i^-g^{ij}r_j^-)}\nn\\&
&\times\,f_{q^++r^+,q^-+r^-}
\label{tachyonaction}\eea
for $f\in{\cal S}(\Gamma^*\oplus\Gamma)$.

The noncommutative gauge theory so constructed in possesses a number of
symmetries. The fact that the action is manifestly gauge invariant leads
immediately to its explicit invariance under target space duality
transformations (represented by the actions of inner automorphisms) and
also its general covariance. It is also invariant under the outer
automorphism group
\eqn{outeralg} representing ``diffeomorphisms'' of the noncommutative geometry.
This includes the Morita equivalence acting on the doublet of vector
potentials by the vector representation of $O(d,d;\zeds)$, and also the
$O(2;\reals)$ chiral rotations of the doublet.
Like the usual
formulations of duality symmetric action functionals, this model possesses
an $O(2;\reals)$ doublet of vector potentials, but the present gauge theory
is generally covariant without the need of introducing auxiliary fields.
The price to pay for this is the nonlocality of the gauge interactions, but
from the point of view of quantum field theoretic renormalizability the
present model is in fact completely well-defined. The gauge theory can be
considered as a dynamical model which measures how much duality symmetry is
present in the target space and hence how far away the stringy perturbation
is from ordinary classical spacetime \cite{lls}. The action functional can
thereby be regarded as an effective measure of distance scales in the
spacetime.

The nonlocality which arises in this field theory can also be shown to be
induced by D1-branes wrapping a highly oblique 2-torus \cite{dh}, whose
worldvolume field theory is thereby described by 2+1 dimensional noncommutative
gauge theory. By $T$-duality, this gauge theory also describes D0-branes on a
very small 2-torus with non-zero Neveu-Schwarz two-form field, which in the
large $N$ limit becomes the Matrix Theory description of M-theory with a
non-zero background three-form field and a light-like compactification circle.
It would be interesting to pursue further the interrelationships that exist
between the two noncommutative geometric approaches, one based on (worldsheet)
vertex operator algebras and the other on (target space) Matrix Theory
compactifications, to string theory and M-theory. Their very natural
relationships to the noncommutative torus hints that the connection may not be
very far away.

\end{document}